\documentclass[seceq]{iis}
\usepackage{graphics}
\usepackage{color}
\usepackage{amsmath}
\usepackage{setspace}
\usepackage{caption}

\subjclass{Primary 62F12, Secondary 62F05.}
\title{A challenge for critical point of spin glass in ground state}
\runtitle{A challenge for critical point of spin glass in ground state}

\author{%
\name{Masayuki \surname{OHZEKI}}
\CAE{mohzeki@i.kyoto-u.ac.jp} }
\inst{Department of Systems Science, Graduate School of Informatics, Kyoto University, \address{Kyoto 606-8501, Japan}}
\runauthor{Masayuki Ohzeki}

\support{This work is supported by MEXT in Japan, Grant-in-Aid for Young Scientists (B) No.24740263.}

\abst{We show several calculations to identify the critical point in the ground state in random spin systems including spin glasses on the basis of the duality analysis.
The duality analysis is a profound method to obtain the precise location of the critical point in finite temperature even for spin glasses.
We propose a single equality for identifying the critical point in the ground state from several speculations.
The equality can indeed give the exact location of the critical points for the bond-dilution Ising model on several lattices and provides insight on further analysis on the ground state in spin glasses.
}

\kword{\kw{Spin Glass}, \kw{Ground state}, \kw{Frustration}, \kw{Percolation}, \kw{Critical point}, \kw{Duality}}

\begin{document}
\maketitle

\section{Introduction}
\label{Introduction}

Spin glass is a typical random system, which spreads over interdisciplinary realms such as information science \cite{HNbook}.
Mean-field analyses have elucidated various aspects of nature of spin glasses in a satisfiable level, although understanding of finite dimensional spin glasses is not yet established.
Recently, successful results for spin glasses have been obtained via the duality analysis \cite{Kramers1941,Wu1976}, which is a standard technique to identify the critical point in finite-dimensional spin systems \cite{Ohzeki2008,Ohzeki2009,Ohzeki2009a,Ohzeki2013a}.
The duality analysis can also identify theoretical limitation of several quantum error correcting codes through their correspondence to the spin glass models \cite{Ohzeki2009b,Ohzeki2012a,Ohzeki2012b}.
The special techniques of the quantum information, not only the error correcting code but also some computation algorithm, are closely related to the spin glass model \cite{Fujii2013}.
Thus the reliable method to investigate the nature of the finite-dimensional spin glass models is desired. 

The duality is known to be a hidden symmetry between the partition functions in low and high temperatures.
This symmetry allows us to identify the locations of the critical points for various spin models without any randomness such as the Ising and Potts models \cite{Kramers1941,Wu1976}.
The straightforward application of the standard duality analysis can not give exact locations of the critical point of spin glasses, but the improved method with the aid of the concept of real-space renormalization group can give the precise estimations in high-temperature regions \cite{Ohzeki2008,Ohzeki2009}.
However we do not yet establish a method to identify the critical point in the low-temperature regions for spin glasses.
In the present study, we revisit this challenging problem by reviewing recent development of the duality analysis for random spin systems including spin glasses from a point of view of frustration, which is an essential nature of emergence of spin-glass behavior.

\section{Duality analysis for spin glasses}
Our analysis is based on the duality \cite{Kramers1941,Wu1976}, which is the simplest way to identify the critical point of the classical spin system.
It is straightforward to generalize the standard duality analysis to the case of spin glasses.
\subsection{Spin glass}
We consider the random-bond Ising model with the following Hamiltonian 
\begin{equation}
H = - \sum_{\langle ij \rangle} J_{ij} S_iS_j,\label{Ham}
\end{equation}
where $S_i$ stands for the Ising spin taking $\pm 1$, and $J_{ij}$ stands for the random coupling assumed to follow various type of a distribution function.
The partition function is
\begin{equation}
Z = \sum_{S_i} \prod_{\langle ij \rangle} \exp(K \tau_{ij} S_iS_j)
\end{equation}
where $K=\beta J$ ($\beta$ is an inverse temperature) and $J_{ij} = J\tau_{ij}$.
In the present study, we deal with $\pm J$ Ising model, which is a typical model of the spin glass, with the following bimodal distribution
\begin{equation}
P(\tau_{ij}) = p \delta(\tau_{ij}-1) + (1-p)\delta(\tau_{ij}+1) = \frac{\exp(K_p \tau_{ij})}{2\cosh K_p},
\end{equation}
where $\exp(-2K_p)=(1-p)/p$.
Although analysis on spin glasses in finite dimensions is an intractable task in general, the special symmetry embedded in a subspace $K=K_p$ known as the Nishimori line enables us to perform the exact/rigrous treatment on several quantities \cite{HNbook,Nishimori1981}.

In random spin systems, we take the configurational average of $J_{ij}$ in order to evaluate the free energy.
The replica method is the most popular technique to perform the configurational average.
Instead of the averaged logarithm of the partition function (free energy), we analyze the averaged power following the well-known identity as
\begin{equation}
\left[ \log Z \right] = \lim_{n \to 0} \frac{\left[Z^n\right]-1}{n},
\end{equation}
where $[\cdots]$ stands for the configurational average.
We then regard the averaged power of the partition function $[Z^n]$ as the effective partition function written as $Z_n$ (the replicated partition function).
At the first stage of analysis, we deal with the replicated system by setting $n$ as a natural number.
At the final step of analysis, we take the limit of $n \to 0$.

\subsection{Standard duality analysis}
\begin{figure}[tbp]
\begin{center}
\includegraphics[width=80mm,keepaspectratio,clip]{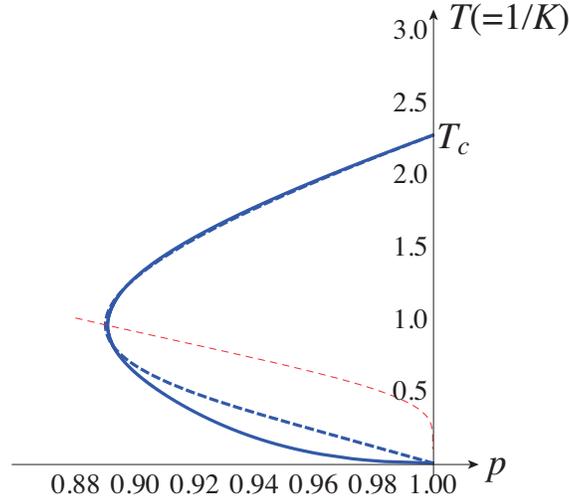}
\end{center}
\caption{Phase diagram given by the duality analysis.
We set $J=1$ for simplicity. 
The blue dashed curve depicts the result from the standard duality analysis.
The red dashed curve denotes the Nishimori line.
The horizontal axis expresses the concentration of the antiferromagnetic interactions ($\tau_{ij}=-1$) and the vertical one represents the temperature.
The critical point of the Ising model on the square lattice without any disorder is located at $T_c$.
For reference, we describe the blue bond curve obtained by the improved method by the aid of the real-space renormalization.}
\label{fig1}
\end{figure}

Let us perform the duality analysis to the effective partition function $Z_n$.
The duality is based on an inherent symmetry embedded in the partition function \cite{Kramers1941}.
Two different approaches to calculate the partition function, the low- and high-temperature expansions, can be related to each other by the $n$-multiple binary Fourier transformation for the local part of the Boltzmann factor, namely edge Boltzmann factor \cite{Wu1976,Nishimori2002,Nishimori2003}.
The effective partition function consists of the following edge Boltzmann factor as
\begin{equation}
x_{\{S^{\alpha}_{i},S_j^{\alpha}\}} = \left[ \prod_{\alpha=1}^n\exp(K \tau_{ij}S^{\alpha}_{i}S^{\alpha}_j) \right].
\end{equation}
The superscript $\alpha$ runs from $1$ to $n$ standing for the index of the replicas.
On the other hand the dual edge Boltzmann factor is obtained by $n$-multiple binary Fourier transformation as
\begin{eqnarray}\nonumber
&&x^*_{\{S^{\alpha}_{i},S_j^{\alpha}\}} = \left(\frac{1}{\sqrt{2}}\right)^n\left[ \prod_{\alpha=1}^n\left({\rm e}^{K \tau_{ij}} + S^{\alpha}_{i}S_j^{\alpha}{\rm e}^{K \tau_{ij}} \right) \right].\\
\end{eqnarray}
Each term in the low-temperature expansion of the effective partition function can be written by $x_{\{S^{\alpha}_i,S^{\alpha}_j\}}$, while the high-temperature one is given by the dual edge Boltzmann factor $x^*_{\{S^{\alpha}_{i},S^{\alpha}_{j}\}}$.
Consequently, we find a double expression of the partition function by use of two different edge Boltzmann factors as
\begin{equation}
Z_n(x_0,x_1,\cdots) = 2^{n(N_S-N_B/2-1)}Z_n^*(x^*_0,x^*_1,\cdots) \label{duality0}.
\end{equation}
where $Z_n^*$ is the partition function on a dual lattice.
Here we simply express each term of $x_{\{S_i^{\alpha},S_j^{\alpha}\}}$ as $x_k$ and $x^*_{\{S_i^{\alpha},S_j^{\alpha}\}}$ as $x^*_k$, where the subscript $k$ denotes the configuration of the edge spins.
In the replicated partition function, $k$ usually describes the number of pairs with edge spins antiparallel among $n$ replicas.  
Here $N_S$ and $N_B$ denote the numbers of sites and plaquettes, respectively.
Notice that $N_B/2=N_S$ on the square lattice.
The unity in the power of $2$ can be ignored in the following analysis.
We obtain another system on the dual graph, on which each site on the original lattice exchanges with each plaquette on the dual one and vice versa, after the so-called dual transformation through the $n$-multiple binary Fourier transformation.
When the dual lattice is the same as the original one, we state that the system holds self-duality.
For instance, the square lattice under consideration is the case.
Then $Z_n^*(x^*_0,x^*_1,\cdots) = Z_n(x^*_0,x^*_1,\cdots)$.
Let us extract the principal Boltzmann factors with edge spins parallel $x_0$ and $x_0^*$ from both sides of Eq. (\ref{duality0}) as
\begin{equation}
(x_0)^{N_B}z_n(u_1,u_2,\cdots) = (x^*_0)^{N_B}z_n(u^*_1,u^*_2,\cdots),
\end{equation}
where $z$ is the normalized partition function $z_n(u_1,u_2,\cdots)=Z_n/(x_0)^{N_B}$ and $z_n(u^*_1,u^*_2,\cdots)=Z_n/(x^*_0)^{N_B}$.
We define the relative Boltzmann factors $u_k = x_k/x_0$ and $u_k^*= x^*_k/x^*_0$.
Nishimori {\it et al.} proposed a conjecture to derive the critical point of the $\pm J$ Ising model on the Nishimori line through the following simple equality \cite{Nishimori2003}
\begin{equation}
x_0 = x_0^*.\label{MCP_duality}
\end{equation}
If we take the limit of $n \to 0$, we can obtain 
\begin{equation}
\left[ \log \left\{ 1 + \exp(-2K\tau_{ij}) \right\} \right] = \frac{1}{2} \log 2.\label{MCPduality}
\end{equation}
On the Nishimori line $K=K_p$, the above equality can be reduced to
\begin{equation}
-p \log p - (1-p) \log (1-p) = \frac{1}{2} \log 2,
\end{equation}
which gives $p_c \approx = 0.8900$.
This estimation is very reasonable and consistent with several results by numerical simulations.
If we remove the restriction on the Nishimori line, we obtain a provisional description of the whole phase diagram by solving the following equality as in Fig. \ref{fig1}
\begin{equation}
p \log ( 1 + e^{-2K} ) + (1-p)\log (1+ e^{2K}) = \frac{1}{2} \log 2.\label{MCPpmJ}
\end{equation}
%-----------------------------------
%-----------------------------------
This equality includes the exact solution of the Ising model without any disorder as $\exp(-2K_c) = \sqrt{2}-1$ for $p=0$ \cite{Kramers1941}.
In addition, the critical point on the Nishimori line is known to be located at the furthest from the $T$ axis \cite{HNbook,Nishimori1981}.
Therefore the phase diagram described by the above equality is approximately correct in the high-temperature region beyond the Nishimori line.

However the duality analysis can not give any $p_c$ on the ground state ($T=0$).
Following the well-known results, the phase boundary is expected to reach a finite value of $p$ in the ground state.
The duality analysis fails to predict the accurate phase boundary in the low-temperature region below Nishimori line.
Recently an improved method with the aid of the concept of real-space renormalization can lead to more precise estimations of the locations of the critical points for spin glasses.
It is however unsuccessful in deriving the finite value of the critical point in  the ground state, although the improved method shows systematic improvement toward the vertical shape around the Nishimori line \cite{Ohzeki2008,Ohzeki2009}. 
Thus we have to add another ingredient to derive the critical point on the ground state.
In the present paper, we show a challenge from several speculations in progress.

\subsection{Ground state via duality analysis}
Let us look at the obtained equality (\ref{MCPpmJ}) by the standard duality analysis.
The inside of the logarithm stems from the ratio of two principal Boltzmann factors.
The original principal Boltzmann factor is $\exp(K \tau_{ij})$ representing the local contribution with edge spins parallel.
On the other hand, the dual principal Boltzmann factor is written as $\propto \exp(K \tau_{ij}) + \exp(-K \tau_{ij})$, which are the summation of the local Boltzmann factor over possible configurations of the edge spins except for the overall inversion.
The right hand side of Eq. (\ref{MCPpmJ}) is related to the number of bonds to be considered here.
If we take the limit of $K \to \infty$, the second term diverges.
Thus we have to assume $p_c$ should vanish as $p_c \sim 1/K$.

On the other hand, for the case of the bond-dilution Ising model, which is one of the typical random spin systems, Eq. (\ref{MCPduality}) can describe the approximately correct phase diagram.
The Hamiltonian is the same form in Eq. (\ref{Ham}), but the distribution function differs from the case of the $\pm J$ Ising model as
\begin{equation}
P(\tau_{ij}) = p \delta(\tau_{ij}-1) + (1-p)\delta(\tau_{ij}).
\end{equation}
Then Eq. (\ref{MCPduality}) is reduced to 
\begin{equation}
p \log ( 1 + e^{-2K} ) + (1-p) \log 2 = \frac{1}{2} \log 2. \label{Per0}
\end{equation}
In the ground state $K \to \infty$, the first term representing the contribution of the ferromagnetic interaction vanishes.
This fact implies the existence of the unique ground state for the single bond.
On the other hand, the logarithm of $2$ on the left hand side expresses the number of degeneracy of the ground state on the single bond with absence of the interaction except for the overall inversion.
We obtain the well-known bond-percolation threshold on the square lattice as $p_c = 1/2$ from Eq. (\ref{Per0}).
We can thus infer the essential term to derive the critical point is yielded from the degeneracy of the ground state, which is an essential property of spin glasses such as frustration.

\section{Triangular lattice: Duality analysis with star-triangle transformation}
In this section, we give several calculations on the critical point on the ground state in a different system.
Let us first consider the bond-dilution Ising model on the triangular lattice.
\subsection{Star-triangle transformation}
\begin{figure}[tbp]
\begin{center}
\includegraphics[width=90mm,keepaspectratio,clip]{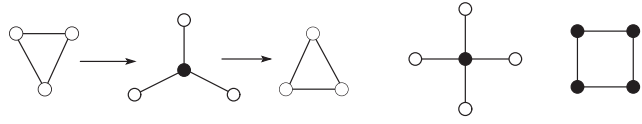}
\end{center}
\caption{The unit cells on the triangular lattice (three of the left panel) and square lattice (two of the right panel).
The black circles denote the internal spins that we sum over, while the white ones are fixed as $S_i = 1$.
The left panel shows the unit cell to be considered on the triangular lattice (down-pointing triangle), the dual hexagonal lattice (up-pointing star), and the dual triangular lattice (up-pointing triangle).
The right panel shows the unit cell for the square lattice.}
\label{fig2}
\end{figure}

If we apply the duality analysis to a system on the triangular lattice, the original triangular lattice is changed into the hexagonal lattice.
We thus have to employ another technique to relate the hexagonal lattice to the original triangular lattice.
This can be achieved by the partial summation over internal spins at the up-pointing star, we call the unit cell below, on the hexagonal lattice, namely star-triangle transformation \cite{Wu1982}.
Then we can transform the partition function on the hexagonal lattice into that on another triangular lattice, namely $Z_n^*(x^*_0,x^*_1,\cdots)=Z_n(x^{*({\rm tr})}_0,x^{*({\rm tr})}_1,\cdots)$ in Eq. (\ref{duality0}).
We here use the renormalized-edge Boltzmann factor $x^{*({\rm tr})}_k$ defined as
\begin{equation}
x_{\{S_i^{\alpha}\}}^{*({\rm tr})} = \left(\frac{1}{\sqrt{2}}\right)^n \left[  \prod_{\alpha} \sum_{S^{\alpha}_0} \prod_{i}\left(\frac{1}{\sqrt{2}}\right)\left(e^{K\tau_{0i}} + S^{\alpha}_0 S^{\alpha}_i e^{-K\tau_{0i}} \right)\right], \label{RBS}
\end{equation}
where the product runs over the three bonds on the unit cell of the hexagonal lattice, namely the up-pointing star.
The central spin is denoted by the subscript of $0$, and the surrounding spins are by $i =1,2$ and $3$.
Here we regard the subscript $k$ as the spin configurations among three sites on the unit cell.
We take the summation over the central spin $S_0$ on the unit cell denoted by the black circle as in Fig. \ref{fig2}.
%-----------------------------------
%-----------------------------------
The coefficient $1/\sqrt{2}$ comes from that in front of the partition function on the right-hand side of Eq. (\ref{duality0}).
Notice that $N_S$ is the same as the number of the unit cells $N_{\rm tr}$ on the triangular lattice, and $N_B=3N_S$.
On the other hand, we rewrite the original partition function in terms of the product of the edge Boltzmann factors as
\begin{equation}
x_{\{S_i^{\alpha}\}}^{({\rm tr})} = \left[ \prod_{\alpha} \exp\left\{K (\tau_{01} S_2^{\alpha}S_3^{\alpha} + \tau_{02} S_3^{\alpha}S_1^{\alpha} + \tau_{03} S_1^{\alpha}S_2^{\alpha}) \right\}\right].\label{RBT}
\end{equation}
Similarly to the case on the square lattice, we put the simple equality to identify the critical point as \cite{Nishimori2006}
\begin{equation}
x^{({\rm tr})}_0 = x^{*({\rm tr})}_0.\label{MCP_th}
\end{equation}
where $x^{({\rm tr})}_0 = x^{({\rm tr})}_{\{S_i^{\alpha}=1\}}$ and $x^{*({\rm tr})}_0 = x^{*({\rm tr})}_{\{S_i^{\alpha}=1\}}$.
This equality again derives the approximately correct locations of the critical points for the $\pm J$ Ising model above the Nishimori line \cite{Nishimori2006}.
The critical points in the whole regions of the bond-dilution Ising model can be also estimated by Eq. (\ref{MCP_th}).  
Let us carefully look at the detailed calculations for two cases below.

\subsection{Critical points in ground state: bond-percolation threshold}
Taking the limit of $n \to 0$, we obtain the following equation from Eq. (\ref{MCP_th})
\begin{equation}
\left[ \log \left( 1 + e^{-2K(\tau_{01}+\tau_{02})}+ e^{-2K(\tau_{02}+\tau_{03})}+ e^{-2K(\tau_{03}+\tau_{01})} \right) \right] = \log 2.
\label{MCP_tr_d}
\end{equation}
The term on the right hand side comes from the geometric features of the triangular lattice.
On the other hand, the inside of the logarithm on the left-hand side is given by the ratio between the two principal Boltzmann factors.
The original principal Boltzmann factor comes from the unit cell with edge spins parallel, and the dual one coincides with the summation over all possible configurations on the dual unit cell except for the overall inversion.

For the bond-dilution Ising model, Eq. (\ref{MCP_tr_d}) can be reduced to 
\begin{equation}
p^3 \log \left( 1 + 3e^{-4K}\right) + 3 p^2(1-p) \log \left( 1 + 2e^{-2K} + e^{-4K}\right) + 3 p(1-p)^2 \log \left( 2 + 2e^{-2K} \right) + (1-p)^3 \log 4 = \log 2.
\end{equation}
In order to estimate the location of the critical point in the ground state, we take the limit of $K \to \infty$ and obtain
\begin{equation}
3 p(1-p)^2 \log 2 + (1-p)^3 \log 4 = \log 2.\label{PerT}
\end{equation}
The first term on the left-hand side consists of the number of the degeneracy of the state on the unit cell of the triangular lattice with two of the three bonds absent, and the second one is similar quantity with none of the three bonds ferromagnetic.
We obtain the bond percolation threshold $p_c = 0.3473$ on the triangular lattice from the following equation to which Eq. (\ref{PerT}) is reduced
\begin{equation}
p^3 -3p +1 = 0.
\end{equation}
If we deal with the inhomogeneous case of the bond-dilution Ising model with different concentrations among three bonds on the unit cell of the triangular lattice as $p$, $r$ and $s$, we can also derive the critical manifold of the bond-percolation threshold (critical points in the ground state) via the same duality analysis as \cite{Ohzeki2013a,Ohzeki2013}
\begin{equation}
prs - p -r -s +1 = 0. \label{Teq}
\end{equation}
Then we obtain a relation consisting of the degeneracy in the ground state similarly to the case of the homogeneous  bond-percolation problem.

\subsection{Critical points in ground state: spin glass}
If we consider the $\pm J$ Ising model, we have several diverging terms, which stems from antiferromagnetic interactions, differently from the case of the bond-dilution Ising model.
Therefore we cannot straightforwardly derive the location of the critical point of spin glasses in the ground state only via the duality analysis.
We must cope with the diverging terms to elucidate relevant results.
If we simply omit the diverging terms in the ground state for the $\pm J$ Ising model, we obtain an interesting consequence.
For the $\pm J$ Ising model on the triangular lattice, Eq. (\ref{MCP_tr_d}) can be reduced to 
\begin{eqnarray}\nonumber
&&p^3 \log ( 1 + 3e^{-4K} ) + 3 p^2(1-p)\log ( 3 + e^{-4K} ) + 3 p(1-p)^2\log ( 1 + 3e^{-4K}) + (1-p)^3 ( 3 + e^{-4K}) \\
&& \quad + 4K \left\{ 3p(1-p)^2 + (1-p)^3\right\} = \log 2.
\end{eqnarray}
If we take the limit of $K \to \infty$ while omitting the diverging term $4K \left\{ 3p(1-p)^2 + (1-p)^3\right\}$, we obtain
\begin{equation}
\{3 p^2(1-p) + (1-p)^3\} \log 3 = \log 2.\label{abv}
\end{equation}
The value satisfying the above equality is estimated as $p_c=0.1801$ and $0.6599 \pm 0.2770{\rm i}$.
The expected value, which is given by numerical simulation, is not close to this estimation as $p_c \approx 0.84$ \cite{Fujii2012}.
However the above equality provides an interesting speculation.
The remaining two terms come from the bond configuration ($\{\tau_{ij}\}$) with frustration, which plays an essential roll in the ground state of spin glasses.
The existence of frustration yields degeneracy in the ground state and generates the number of the inside of the logarithm.
Therefore we find that the above equality (\ref{abv}) consists of the number of degeneracy in the ground state similarly to the case of the bond-dilution Ising model by simply removing the diverging terms.

\section{Square lattice: duality analysis with real-space renormalization}
The standard duality analysis can not give the exact solution in the random spin systems.
However if we introduce the partial summation after the standard duality analysis inspired by real-space renormalization, we improve the precision of the estimation on the location of the critical point.

\subsection{Duality analysis with real-space renormalization}
We start from Eq. (\ref{duality0}) for the case on the square lattice.
Notice that the edge Boltzmann factor is not enough to express the inherent nature of the random spin systems such as frustration in spin glasses.
Thus we consider to use the renormalized-edge Boltzmann factor inspired by the star-triangle transformation.
We take a unit cell consisting of four bonds from both of the original and dual square lattices as in Fig. \ref{fig2}.

Let us take the product of the edge Boltzmann factors and perform the summation over the internal spin.
In order to evaluate the renormalized-edge Boltzmann factors, we fix the edge spins to $S_i=1$ on the cluster and sum over the internal spin similarly to the star-triangle transformation.
The renormalized-principal Boltzmann factor is written as 
\begin{equation}
x_0^{({\rm sq})} = \left[ \left\{ \sum_{S_0} \prod_{i}\exp(K \tau_{0i}S_0) \right\}^n\right],
\end{equation}
where the product runs over four bonds on the unit cell.
The dual renormalized-principal Boltzmann factor is given by 
\begin{equation}
x_0^{*({\rm sq})} = \left[ \left\{ \left(\frac{1 }{\sqrt{2}}\right)^4\sum_{S_0} \prod_{i}\left({\rm e}^{K \tau_{0i}}+{\rm e}^{-K \tau_{0i}}S_0 \right) \right\}^n\right].
\end{equation}
This quantity corresponds to a local partition function for the dual unit cell consisting of the single square.
Similarly, we impose the following equation to identify the location of the critical point
\begin{equation}
x^{({\rm sq})}_0 = x^{*({\rm sq})}_0.\label{MCP1}
\end{equation}
Taking $n \to 0$ in Eq. (\ref{MCP1}), we obtain the following formula
\begin{eqnarray}\nonumber
&&\left[ \log \left( \frac{ \prod_{i}2\cosh K \tau_{0i}}{2\cosh \sum_i K \tau_{0i}}  \right) \left(1 + \prod_i\tanh K \tau_{0i}\right)\right] = 2 \log 2. \label{PP}\\
\end{eqnarray}
The term on the right hand side comes from the geometric features of the unit cluster with four bonds.
On the other hand, the inside of the logarithm on the left-hand side is given by the ratio between the two principal Boltzmann factors.
The original principal Boltzmann factor comes from the unit cluster with edge spins parallel while the center spin is summed over, and the dual one coincides with the summation over all possible configurations on the dual unit square except for the overall inversion.

If we estimate the locations of the critical points in the whole regions, we confirm their precision to be enhanced by the real-space renormalization for both of the bond-dilution Ising model and $\pm J$ Ising model.
For the bond-dilution Ising model, we can reproduce the bond-percolation threshold in the ground state $K \to \infty$ as $p_c = 1/2$ as well as the improvement of estimations in the whole regions.
By taking the limit of $K \to \infty$, we obtain the following equation
\begin{eqnarray}\nonumber
&& p^4 \log 2 + 4p^3(1-p) \log 2 + 6p^2(1-p)^2 \log 4 + 4p(1-p)^3 \log 8 + (1-p)^4 \log 8 = 2 \log 2.\label{PPbefore}
\end{eqnarray}
By recalling a simple identity $p^4 + 4p^3(1-p) + 6p^2(1-p)^2 + 4p(1-p)^3 + (1-p)^4 =1$, we obtain 
\begin{eqnarray}\nonumber
6p^2(1-p)^2 \log 2 + 4p(1-p)^3 \log 4 + (1-p)^4 \log 4 = \log 2.
\end{eqnarray}
Each term of the inside of the logarithm consists of the number of degeneracy without the overall inversion of spins for each bond configurations $\{\tau_{0i}\}$.
The solution indeed becomes $p_c = 1/2$.
In addition, even if we deal with the inhomogeneous case of the bond-dilution Ising model with the different probabilities as $p$, $r$, $s$ and $t$ on each unit cell, we can obtain the exact critical manifold of the bond-percolation thresholds,
\begin{eqnarray}
prs +prt+rst+pst -pr -ps -rs -pt -rt -st +1 = 0.\label{Ceq}
\end{eqnarray}

\subsection{Ground state of spin glass on the square lattice}
Let us consider the case of the $\pm J$ Ising model on the square lattice.
Equation (\ref{PP}) is written as
\begin{eqnarray}\nonumber
&&\{p^4 + (1-p)^4\} \log \left( \frac{2 + 12e^{-4K}+ 2 e^{-8K}}{1 + e^{-8K}}\right) + 6 p^2(1-p)^2 \log \left( \frac{ 2e^{4K} + 12 + 2 e^{-4K}}{2}\right) \\ 
&&\quad + 4\{ p^3(1-p)+ p(1-p)^3\} \log \left( \frac{ 8 + 8 e^{-4K}}{1 + e^{-4K}}\right) = 2 \log 2.
\end{eqnarray}
We simplify the above equation into, by extracting the diverging term as $24 K p^2(1-p)^2$,
\begin{eqnarray}
&& - 6 p^2(1-p)^2 \log 2 + 4\{ p^3(1-p)+ p(1-p)^3\} \log 4 = \log 2.
\end{eqnarray}
The reason why the first term becomes negative stems from the degeneracy of the ground state of the original unit cell with two of the interactions antiferromagnetic.
Thus we can infer the generic equation obtained from the duality analysis with real-space renormalization as
\begin{equation}
\left[ \log D^*_{\rm GS}(\{J_{ij}\}) \right] - \left[ \log D_{\rm GS}(\{J_{ij}\}) \right] = g \log 2.\label{conjecture}
\end{equation}
where $D_{\rm GS}$ and $D^*_{\rm GS}$ are the number of degeneracy in the ground state for the original and dual unit cells under consideration, respectively.
In addition, $g$ is a constant depending on the geometric property of the lattice such that $g=1$ for the case of the triangular unit and for that of the square lattice.
However we cannot find any relevant real-valued estimation in the case with four-bond unit.
The solutions are $0.8152 \pm 0.1766{\rm i}$ and $0.1848 \pm 0.1766{\rm i}$.
Thus we do not reach the precise value of the critical point in the ground state.

Unfortunately we can not yet obtain the relevant results.
The duality analysis fully exploits the particular symmetry hidden in two-dimensional systems.
Therefore we may expect that the duality analysis, which has been available in finite temperature, can open the way to elucidate the property in the ground state of spin glasses.
The simple reduction of the diverging terms would be problematic but can yield the degeneracy in the given bond configuration, which is the most relevant in the ground state.
The diverging terms stemming from difference of the ground-state energy for the original and dual unit cells.
Therefore, if we prepare the adequate unit cell not to generate any difference of the ground state energy, we might find relevant results in the ground state of spin glasses.
In addition, we may increase the degree of the real-space renormalization to reach the relevant result such that the duality analysis with real-space renormalization can lead to the exact solution of the phase boundary by use of the large unit cell \cite{Ohzeki2011}.
The complex number of the solutions might imply the existence of the possible way to investigate the critical property in the ground state such as the partition-function zeros \cite{HNbook2}.
In infinite number of degrees of freedom, the partition function zeros reach the real-value axis and show the exact solution of the critical point.

\section{Summary}
We show several speculations on the critical point of spin glasses via the duality analysis and its improved method with real-space renormalization.
We did not obtain any relevant results on the main problem related to the ground state of the spin glass, but the method we employed in the present study is based on the exact theory applicable to the bond-dilution Ising model.
As shown in the same issue of this proceedings, Miyazaki recently gave an interesting theory to estimate the approximate location of the critical point in the ground state of spin glasses.
His theory is based on the number of the frustration emerging in the system but its theoretical validity is still a little bit ambiguous.
Our study shown in this article is similarly related to the frustration but its number and degeneracy involved in its existence in the ground state, and our theory stems from the duality analysis.
We hope that further study connects two of the newborn theories and solves intractable problems to investigate the property of the ground state of spin glasses in an analytical way.

\end{document}